\documentclass[twocolumn,showpacs,aps,prl,superscriptaddress,letterpaper]{revtex4}
\usepackage{graphicx}
\usepackage{dcolumn}
\usepackage{amsmath}
\usepackage{epsfig}
\usepackage{multirow}
\usepackage{subfigure}

\long\def\inst#1{\par\nobreak\kern 4pt\nobreak
    {\itshape #1}\par\vskip 10pt plus 3pt minus 3pt}
\RequirePackage{xspace}
\usepackage{relsize}

\newcommand{\BABARPubYear}    {10}
\newcommand{\BABARPubNumber}  {23}

\newcommand{\SLACPubNumber} {14204}
\newcommand{\LANLNumber} {1007.4646}

\input babarsym
\def\cpoddhiggs      {\ensuremath{{A^{0}}}\xspace}

\def\dm          {\ensuremath{\chi}\xspace}
\def\dmbar       {\ensuremath{{\overline\chi}}\xspace}

\def\n1Spipi     {\ensuremath{}\xspace}

\def\beq{\begin{equation}}
\def\eeq{\end{equation}}
\def\bea{\begin{eqnarray}}
\def\eea{\end{eqnarray}}
\def\bq{\begin{quote}}
\def\eq{\end{quote}}
\def\bi{\begin{itemize}}
\def\ei{\end{itemize}}
\def\bc{\begin{center}}
\def\ec{\end{center}}

\def\etal{{\em et al.}}

\def\dm          {\ensuremath{\chi}\xspace}
\def\dmbar       {\ensuremath{{\overline\chi}}\xspace}
\def\invisible   {\ensuremath{\mathrm{invisible}}\xspace}

\newcommand{\higgsmass}{\ensuremath{m_{A^0}}\xspace}
\newcommand{\dmmass}{\ensuremath{m_{\dm}}\xspace}

\newcommand{\recMass}{\ensuremath{M_\mathrm{recoil}}\xspace}
\newcommand{\missMass}{\ensuremath{M_X^2}\xspace}
\def\ppbar      {\ensuremath{p\overline{p}}\xspace}
\def\nnbar      {\ensuremath{n\overline{n}}\xspace}

\begin{document}

\onecolumngrid
\hbox to \hsize{
\vbox{

\begin{flushleft}
%% \babar\ Analysis Document \# 2330, Version 12\\
July 2010 \\
\end{flushleft}
%\vspace{\baselineskip}
\vspace{\baselineskip}
}
\hfill
\vbox{
\begin{flushright}
\babar-PUB-\BABARPubYear/\BABARPubNumber \\
SLAC-PUB-\SLACPubNumber \\
arXiv:\LANLNumber [hep-ex] \\
\end{flushright}
}}
\vspace{-\baselineskip}
\twocolumngrid
% Title of the paper
\title{
\large \bfseries \boldmath
Search for Production of Invisible Final States in Single-Photon 
Decays of $\Upsilon(1S)$
}

%%%%%%%%%%%%%%%%%%%%%%%%%%%%%%%%%%%%%%%%%%%%%%%%%%%%%%%%%%%%%%%%%%%%%%%%%%%%%%%%%%
%% author list :
%
%% author list as of 08-Jul-2010 (435 authors)
%
\author{P.~del~Amo~Sanchez}
\author{J.~P.~Lees}
\author{V.~Poireau}
\author{E.~Prencipe}
\author{V.~Tisserand}
\affiliation{Laboratoire d'Annecy-le-Vieux de Physique des Particules (LAPP), Universit\'e de Savoie, CNRS/IN2P3,  F-74941 Annecy-Le-Vieux, France}
\author{J.~Garra~Tico}
\author{E.~Grauges}
\affiliation{Universitat de Barcelona, Facultat de Fisica, Departament ECM, E-08028 Barcelona, Spain }
\author{M.~Martinelli$^{ab}$}
\author{D.~A.~Milanes$^{ab}$}
\author{A.~Palano$^{ab}$ }
\author{M.~Pappagallo$^{ab}$ }
\affiliation{INFN Sezione di Bari$^{a}$; Dipartimento di Fisica, Universit\`a di Bari$^{b}$, I-70126 Bari, Italy }
% ...
\collaboration{The \babar\ Collaboration}
\noaffiliation

%%%%%%%%%%%%%%%%%%%%%%%%%%%%%%%%%%%%%%%%%%%%%%%%%%%%%%%%%%%%%%%%%%%%%%%%%%%%%%%%%%
%                             Abstract                                          %%
%%%%%%%%%%%%%%%%%%%%%%%%%%%%%%%%%%%%%%%%%%%%%%%%%%%%%%%%%%%%%%%%%%%%%%%%%%%%%%%%%%

\begin{abstract}
We search for single-photon decays
of the $\Upsilon(1S)$ resonance, 
$\Upsilon\to\gamma+\invisible$, 
where the invisible state is either a particle of definite mass,
such as a light Higgs boson $\cpoddhiggs$, or
a pair of dark matter particles,
$\dm\dmbar$. Both \cpoddhiggs\ and \dm\ are assumed to have
zero spin. 
We tag \Y1S
decays with a dipion transition $\Y2S\to\pi^+\pi^-\Y1S$ and look for
events with a single energetic photon and significant missing energy. 
We find no evidence for such processes in the mass range 
$\higgsmass\le9.2$\,GeV and $\dmmass\le4.5$\,GeV
in the sample of $98\times10^6$ \Y2S 
decays collected with the \babar\ detector
and set stringent limits on new physics models that contain light
dark matter states. 
\end{abstract}

% PACS, the Physics and Astronomy Classification Scheme.

\maketitle

% The body of the paper starts here
%%%%%%%%%%%%%%%%%%%%%%%%%%%%%%
% INTRODUCTION
%%%%%%%%%%%%%%%%%%%%%%%%%%%%%%

There is compelling astrophysical evidence for the existence of dark
matter~\cite{ref:DM, ref:DMreview}, which amounts to about one quarter
of the total energy density in the Universe. Yet, there is no experimental
information on the particle composition of dark
matter~\cite{ref:DMreview,PDBook}.  
A class of new physics models~\cite{Gunion:2005rw}, motivated by 
astro-particle observations~\cite{INTEGRAL,DAMA},
predicts a light component of the
dark matter spectrum. 
The bottomonium system of $\Upsilon$ states is an ideal
environment to explore 
these models. Transitions 
$\Y3S\to\pi^+\pi^-\Y1S$ and $\Y2S\to\pi^+\pi^-\Y1S$ offer a way to
cleanly detect the production of \Y1S mesons, and enable searches
for invisible or 
nearly invisible decays of the \Y1S~\cite{BAD2009prl}. Such decays
would be a tell-tale sign of low-mass, weakly-interacting dark matter
particles.

The
Standard Model process $\Y1S\to\gamma\nu\bar{\nu}$ is not observable
at the present experimental
sensitivity~\cite{Yeghiyan:2009}. 
An observation of $\Upsilon$ decays
with significant missing energy would be a sign of new physics,
and could shed light on the spectrum of dark matter particles \dm. 
The branching fraction (BF)  
$\BR(\Y1S\to\dm\dmbar)$
is estimated to be as large as
(4--18)$\times10^{-4}$~\cite{McElrath:2007sa,Yeghiyan:2009}, 
while
$\BR(\Y1S \to\gamma\dm\dmbar)$ is
suppressed by $\mathcal{O}(\alpha)$, and the
range $10^{-5}$--$10^{-4}$ is expected~\cite{Yeghiyan:2009}. 

The decays $\Y1S\to\gamma+\invisible$ might also proceed through
Wilczek production~\cite{Wilczek:1977zn} of an on-shell
scalar state \cpoddhiggs: $\Y1S\to\gamma\cpoddhiggs$,
$\cpoddhiggs\to\invisible$. 
Such low-mass Higgs states appear in several extensions of the Standard
Model~\cite{Dermisek:2005ar}. 
Constraining the
low-mass Higgs sector is important for understanding the Higgs discovery
reach of high-energy colliders~\cite{LisantiWacker2009}. 
The BF for $\Y1S\to\gamma\cpoddhiggs$ is predicted to be 
as large as $5\times10^{-4}$, depending on \higgsmass\ and
couplings~\cite{Dermisek:2006py}. If there is also a 
low-mass neutralino with mass $m_{\dm}<\higgsmass/2$, the decays
of \cpoddhiggs\ would be predominantly invisible~\cite{ShrockSuzuki1982}. 

For multibody $\Y1S\to\gamma\dm\dmbar$ decays, the current 90\%
confidence level (C.L.) BF upper limit, based on a data sample of
$\sim10^6$ \Y1S decays, is of order $10^{-3}$~\cite{CLEO:1994ch}.
The limit on two-body $\Y1S\to\gamma+X,\ X\to\invisible$ decays is 
$\mathcal{B}(\Upsilon(1S)\to\gamma+X)<3\times10^{-5}$ for
$m_X<7.2\gev$~\cite{PDBook}. 
The limit on invisible decays of \Y1S is 
$\BR(\Y1S\to\dm\dmbar)<3.0\times10^{-4}$~\cite{BAD2009prl}.

This Letter describes a high-statistics, low-background search for decays 
$\Y1S\to\gamma+\invisible$, characterized by a single
energetic photon and a large amount of missing energy and momentum. 
This is the first search of this kind to use 
the \Y1S mesons produced in dipion 
$\Y2S\to\pi^+\pi^-\Y1S$ transitions. 
We search for both resonant two-body decays 
$\Y1S\to\gamma\cpoddhiggs$, $\cpoddhiggs\to\invisible$, and
nonresonant three-body processes $\Y1S\to\gamma\dm\dmbar$. For the resonant
process, we assume that the decay width of 
the \cpoddhiggs\ resonance is negligible 
compared to the experimental resolution~\cite{ref:Lozano}.
We further assume that both the \cpoddhiggs\ and \dm\ particles have zero
spin. The decays $\Y1S\to\gamma\dm\dmbar$ are modeled with 
phase-space energy and angular distributions, which corresponds to
S-wave coupling between the $b\overline{b}$ and $\dm\dmbar$. 

%%%%%%%%%%%%%%%%%%%%%%%%%%%%%%
%THE \babar\ DETECTOR AND DATASET
%%%%%%%%%%%%%%%%%%%%%%%%%%%%%%

The analysis is based on a sample corresponding to an
integrated luminosity of $14.4\,\mathrm{fb}^{-1}$ 
collected on the \Y2S resonance 
with the \babar\ detector 
at the \pep2\ asymmetric-energy \epem\ collider at the SLAC National
Accelerator Laboratory. This sample corresponds to 
 $(98.3\pm0.9)\times10^6$ \Y2S decays. 
We also employ a sample of $28\,\mathrm{fb}^{-1}$
accumulated on the \Y3S resonance (\Y3S sample) for studies of the continuum
backgrounds. Both $\Y3S\to\pi^+\pi^-\Y2S$ and
$\Y3S\to\pi^+\pi^-\Y1S$ decays produce a dipion system that is
kinematically distinct from the  $\Y2S\to\pi^+\pi^-\Y1S$
transition. Hence, the \Y3S events passing our
selection form a pure high-statistics continuum QED sample. 
For selection
optimization, we also use 
$1.4\,\mathrm{fb}^{-1}$ and $2.4\,\mathrm{fb}^{-1}$ datasets collected
about $30$\,MeV
below the \Y2S and \Y3S resonances, respectively (off-peak samples).
The \babar\ detector, including the tracking and particle
identification systems, the
electromagnetic calorimeter (EMC), and the Instrumented Flux Return
(IFR), is described in detail
elsewhere~\cite{detector,LST}. 

%%%%%%%%%%%%%%%%%%%%%%%%%%%%%%
% EVENT SELECTION
%%%%%%%%%%%%%%%%%%%%%%%%%%%%%%

Detection of low-multiplicity events requires 
dedicated trigger and filter lines. 
First, the hardware-based Level-1 (L1) trigger accepts
single-photon events
if they contain at least one EMC cluster with energy above $800$~MeV.
A collection of L1 trigger patterns based on drift chamber information
selects a pair of low-momentum pions.  
Second, a
software-based Level-3 (L3) trigger 
accepts events with a single EMC cluster with the center-of-mass
(CM) energy $E^{*}_\gamma>1$~GeV~\cite{CMfootnote},
if there is no charged track with transverse momentum 
$p_T>0.25$~GeV originating from the $e^+e^-$ interaction region. 
Complementary to this, a track-based L3 trigger 
accepts events that have at least one track with $p_T>0.2$~GeV. 
Third, an offline filter accepts events that have
exactly one photon with energy $E_\gamma^{*}>1$~GeV, and no tracks 
with momentum $p^{*}>0.5$~GeV. A nearly independent filter
accepts events with two tracks of opposite charge, which form a dipion
candidate with recoil mass (defined below) between $9.35$ and $9.60$
GeV. 

The analysis in the low-mass region
$\higgsmass\le8$~GeV ($m_{\dm}\le4$~GeV), which corresponds to 
photon energies $E_\gamma^{*}>1.1$~GeV, requires
the single-photon or the dipion trigger/filter selection to be satisfied; the
trigger/filter efficiency for signal is $83\%$. In the
high-mass region, $7.5\le\higgsmass\le9.2$~GeV 
($3.5\le m_{\dm}\le4.5$~GeV), 
we only accept events selected with the dipion trigger/filter,  
since a significant fraction of this region lies
below the energy threshold for the single-photon selection.
This
selection has an efficiency of $12.5\%$ for signal events.

We select events with exactly two
oppositely-charged tracks and a 
single energetic photon with 
$E^{*}_\gamma\ge0.15$\,\gev in the central part of the EMC
($-0.73<\cos\theta_\gamma^{*}<0.68$). Additional
photons with $E^{*}_\gamma\le0.12$\,\gev can be present so long as
their summed laboratory energy is less than $0.14$\,\gev. 
We
require that both pions be positively identified with 
85--98\% efficiency for real pions, 
and a misidentification rate of $<5\%$ for low-momentum electrons and $<1\%$
for kaons and protons. 
The pion candidates are required to form a  vertex with
$\chi^2_\mathrm{vtx}<20$ (1 degree of freedom)
displaced in the transverse plane by at most 
2~mm from the 
$e^+e^-$ interaction 
region. The transverse momentum of the pion pair is required to satisfy
$p_{T\pi\pi}<0.5$~GeV, and we reject events if any track has
$p^{*}>1$~GeV. 

We further reduce the background by combining several kinematic
variables of the dipion system~\cite{BAD2009prl} into a multilayer
perceptron neural 
network discriminant (NN)~\cite{TMVA}. 
The NN is trained with
a sample of simulated signal events $\Y1S\to\gamma\dm\dmbar$
($m_{\dm}=0$) and an off-peak sample for background; the NN assigns a
value $\mathcal{N}$ close to $+1$ for signal and close to $-1$ for background. 
We require
$\mathcal{N}>0.65$ in the low-mass region. This selection has an efficiency of 87\% for
signal and rejects 96\% of the continuum background. In the high-mass
region we require $\mathcal{N}>0.89$ (73\% signal efficiency, 98\% continuum
rejection). 

Two additional requirements are applied to reduce specific background
contributions. Neutral hadrons from the radiative decays
$\Y1S\to\gamma\KL\KL$ and $\Y1S\to\gamma\nnbar$ 
may not be detected
in the EMC. 
We remove 90\% of these background events 
by requiring that there be no IFR cluster 
within a range of $20^{\circ}$ of azimuthal angle ($\phi$)
opposite the primary photon (IFR veto).  
This selection is applied for $\higgsmass<4$\,\gev and
$m_{\dm}<2$\,\gev, since the hadronic final states in radiative \Y1S
decays are observed to have low invariant mass~\cite{CLEOgammah+h-}. 

For the high-mass range we
suppress contamination from electron bremsstrahlung by rejecting
events if the photon and one of the tracks are closer than $14^{\circ}$
in $\phi$. In addition, the two-photon process 
$e^+e^-\to e^+e^-\gamma^{*}\gamma^{*}\to e^+e^-\eta'$,
$\eta'\to\gamma\pi^+\pi^-$, in which the $e^+e^-$ pair escapes detection
along the beam axis and the two pions satisfy our selection criteria,
produces photons in a narrow energy range
$0.25<E_\gamma^{*}<0.45$\,\gev. We take advantage of the small transverse
momentum of the $\eta'$ and reject over half of these events by
requiring the primary photon and dipion system to be separated
by at most $\Delta\phi=160^{\circ}$. The signal efficiency for this
requirement is $88\%$. 

The selection criteria are chosen to maximize
$\varepsilon/(1.5+\sqrt{B})$~\cite{Punzi}, 
where $\varepsilon$ is the selection efficiency for 
$\dmmass=0$ and $B$ is the expected background yield. 
The signal efficiency varies between 2 and 11\%, and 
is lowest at the highest masses (lowest photon energy). 
The backgrounds can be classified into three categories: continuum
backgrounds from QED processes $e^+e^-\to\gamma\pi^+\pi^-+\ldots$ with
particles escaping detection, 
radiative leptonic
decays $\Y1S\to\gamma\ell^+\ell^-$,
where leptons $\ell\equiv e,\mu,\tau$ are not detected, and peaking
backgrounds from radiative hadronic decays and two-photon $\eta'$
production. 

%%%%%%%%%%%%%%%%%%%%%%%%%%%%%%
% EXTRACTION OF SIGNAL YIELDS
%%%%%%%%%%%%%%%%%%%%%%%%%%%%%%

We extract the yield of signal events as a function of 
\higgsmass\ ($m_{\dm}$) in the interval $0\le\higgsmass\le9.2$\,\gev\
($0\le m_{\dm}\le4.5$\,\gev)
by performing a series of unbinned extended maximum likelihood scans in steps
of \higgsmass\ ($m_{\dm}$). 
We use two kinematic variables: the dipion recoil
mass \recMass\ and the missing mass squared \missMass:
\begin{eqnarray}
\recMass^{2} &=& M_{\Y2S}^2 + m_{\pi\pi}^{2} - 2M_{\Y2S}E^{*}_{\pi\pi}
\label{eq:Mrecoil}\\
\missMass &=& (\mathcal{P}_{e^{+}e^{-}} - \mathcal{P}_{\pi\pi} -
\mathcal{P}_{\gamma})^{2} 
\label{eq:Mmiss}
\end{eqnarray}
where $E^{*}_{\pi\pi}$ is the CM energy of the dipion system, and
$\mathcal{P}$ is the four-momentum.  
The two-dimensional likelihood function is computed for observables
$(\recMass,\missMass)$ 
over the range
$9.44\le\recMass\le9.48\,\gev$ and
$-10\le\missMass\le68\,\gev^2$ (low-mass region) and 
$40\le\missMass\le84.5\,\gev^2$ (high-mass region). 
It contains contributions from signal, continuum
background, radiative leptonic \Y1S background, 
and peaking backgrounds, as described
below. 
We search for the \cpoddhiggs\ in
mass steps equivalent to half the mass resolution $\sigma(\higgsmass)$.
We sample a total of 196 points in the low-mass
$0\le\higgsmass\le8$\,\gev range, and 146 points in the high-mass range 
$7.5\le\higgsmass\le9.2$\,\gev. For the $\Y1S\to\gamma\dm\dmbar$ search, we
use 17 values of $m_{\dm}$ over $0\le m_{\dm}\le4.5$\,\gev. 
For each \higgsmass\ (\dmmass) value, we compute the value of the negative
log-likelihood $\mathrm{NLL}=-\ln\mathcal{L}(N_\mathrm{sig})$ in steps
of the signal yield $N_\mathrm{sig}\ge0$ while minimizing NLL with
respect to the background yields $N_\mathrm{cont}$ (continuum),
$N_\mathrm{lept}$ ($\Y1S\to\gamma\ell^+\ell^-$), and, where
appropriate, $N_\mathrm{hadr}$ (radiative hadronic background) or
$N_{\eta'}$ (two-photon $\eta'$ background). If the minimum of NLL
occurs for $N_\mathrm{sig}>0$, we compute the raw statistical
significance of a particular fit as
$\mathcal{S}=\sqrt{2\log(\mathcal{L}/\mathcal{L}_0)}$, where 
$\mathcal{L}_0$ is the value of the likelihood for $N_\mathrm{sig}=0$.
For small $\mathcal{S}$, we integrate $\mathcal{L}(N_\mathrm{sig})$
with uniform prior over $N_\mathrm{sig}\ge0$ to 
compute the 90\% C.L. Bayesian upper limits.
In the range
$7.5\le\higgsmass\le8$~GeV and $3.5\le\dmmass\le4$~GeV where the
low-mass and high-mass selections overlap, we add NLLs
from both datasets, ignoring a small (3\%) correlation. 
This likelihood scan procedure is designed to handle samples with a
very small number of events in the signal region. 

We use signal Monte Carlo (MC) samples~\cite{geant,EvtGen}
$\Y1S\to\gamma\cpoddhiggs$ and 
$\Y1S\to\gamma\dm\dmbar$
generated at 17 values of \higgsmass over 
a broad range $0\le\higgsmass\le9.2$\,\gev and at 17 values of $m_{\dm}$
over $0\le m_{\dm}\le4.5$\,\gev 
to determine the signal distributions in \missMass\ and selection
efficiencies. We then interpolate these distributions and efficiencies.
The signal probability density function
(PDF) in \missMass\ is described by a Crystal Ball (CB)
function~\cite{ref:CBshape} ($\Y1S\to\gamma\cpoddhiggs$) or a
resolution-smeared phase-space function ($\Y1S\to\gamma\dm\dmbar$). 
The resolution in \missMass\ is dominated by the photon energy
resolution, and varies monotonically from $1\,\gev^2$ at low
\higgsmass\ to $0.2\,\gev^2$ at $\higgsmass=9.2\,\gev$. 
We correct the signal PDF in \missMass\ for the difference
between the photon energy resolution function in
data and simulation using a high-statistics
$e^+e^-\to\gamma\gamma$ sample. 
We determine the signal 
distribution in \recMass, as well as that
of background containing real \Y1S decays, from a large
data sample of events $\Y1S\to\mu^+\mu^-$.  This PDF is modeled as a
sum of two CB 
functions with common mean, a common resolution
$\sigma(\recMass)\approx2$\,MeV, and two opposite-side tails.

We describe the \missMass\ PDF of the radiative
$\Y1S\to\gamma\ell^+\ell^-$ background by an exponential function, and
determine the exponent from a fit to the 
distribution of \missMass\ in 
a $\Y1S\to\gamma\ell^+\ell^-$ data sample 
in which the two stable leptons ($e$
or $\mu$) are fully reconstructed. Before the fit, this sample is
re-weighted by the probability as a function of
\missMass\ that neither lepton is observed. 

The continuum \missMass\ PDF is described by a function that has a
resolution-smeared phase-space component at low \missMass, and an
exponential rise at high \missMass. For the low-mass selection
($-10\le\missMass\le68\,\gev^2$), we determine
this PDF from a fit to the \Y3S data sample. For the high-mass
region ($40\le\missMass\le84.5\,\gev^2$), we determine this PDF, as well as
the \missMass\ PDF of the peaking $\eta'$ background, from a fit to
the \Y2S data sample selected with the NN requirement $\mathcal{N}<0$.
The \recMass\ PDF is
determined from a fit to the \Y3S data sample. 

The contribution from the radiative hadronic backgrounds is estimated
from the measurement of $\Y1S\to\gamma h^+h^-$
spectra~\cite{CLEOgammah+h-}. We assume isospin symmetry to relate 
$\BR(\Y1S\to\gamma K^+K^-)$ to $\BR(\Y1S\to\gamma\KL\KL)$, and
$\BR(\Y1S\to\gamma\ppbar)$ to $\BR(\Y1S\to\gamma\nnbar)$. A small additional
contribution arises from  $\Y1S\to\gamma\pi^+\pi^-$ events in which
the pions escape detection. 
We expect $N_\mathrm{hadr}=6.6\pm1.1$
radiative hadronic events (without IFR veto), dominated by
$\Y1S\to\gamma\KL\KL$, or 
$N_\mathrm{hadr}^\mathrm{veto}=1.02\pm0.14$ events (with IFR veto). We describe the
\missMass\ distribution of these 
events with a combination of CB functions, using the measured
spectrum of $\Y1S\to\gamma h^+h^-$ events~\cite{CLEOgammah+h-}. 

%%%%%%%%%%%%%%%%%%%%%%%%%%%%%%%%%%%%%%%%
% SYSTEMATIC UNCERTAINTIES
%%%%%%%%%%%%%%%%%%%%%%%%%%%%%%%%%%%%%%%%

The largest systematic uncertainty is on the reconstruction
efficiency, which includes the trigger/filter efficiency
($\varepsilon_\mathrm{trig}$), and photon ($\varepsilon_\gamma$) and dipion
($\varepsilon_{\pi\pi}$) reconstruction and selection efficiencies. We
measure the  product $\varepsilon_{\pi\pi}\times N_{\Y1S}$, where $N_{\Y1S}$ is
the number of produced \Y1S mesons,  
with a clean high-statistics sample of
the $\Y1S\to\mu^+\mu^-$ decays. The
uncertainty ($2.1\%$) is dominated
by $\BR(\Y1S\to\mu^+\mu^-)$ (2\%)~\cite{PDBook} and a small selection
uncertainty for the $\mu^+\mu^-$ final state. We measure 
$\varepsilon_\gamma$ in an
$e^+e^-\to\gamma\gamma$ sample in which one of the photons converts
into an $e^+e^-$ pair in the detector material ($1.8\%$
uncertainty). The trigger efficiency $\varepsilon_\mathrm{trig}$ is
measured in unbiased random samples of events that bypass the
trigger/filter selection. This uncertainty is small for the
single-photon triggers ($0.4\%$), but is statistically limited for the
dipion triggers ($8\%$). In the low-mass region, we take into account
the anti-correlation between single-photon and dipion trigger efficiencies
in L3; the uncertainty for the combination of the 
triggers is $1.2\%$. 

We account for additional uncertainties associated with the
signal and background PDFs, and the predicted number of radiative
hadronic events $N_\mathrm{hadr}$, including PDF parameter
correlations. These uncertainties do not scale with the signal yield,
but are found to be small. 
We also test for possible biases in the fitted value of the signal yield with a
large ensemble of pseudo-experiments. 
The biases are
consistent with zero for all values of \higgsmass\ and \dmmass, and we
assign an uncertainty of $0.25$ events.

%%%%%%%%%%%%%%%%%%%%%%%%%%%%%%
% STATISTICAL INTERPRETATION %
%%%%%%%%%%%%%%%%%%%%%%%%%%%%%%

\begin{figure}
\bc
\epsfig{file=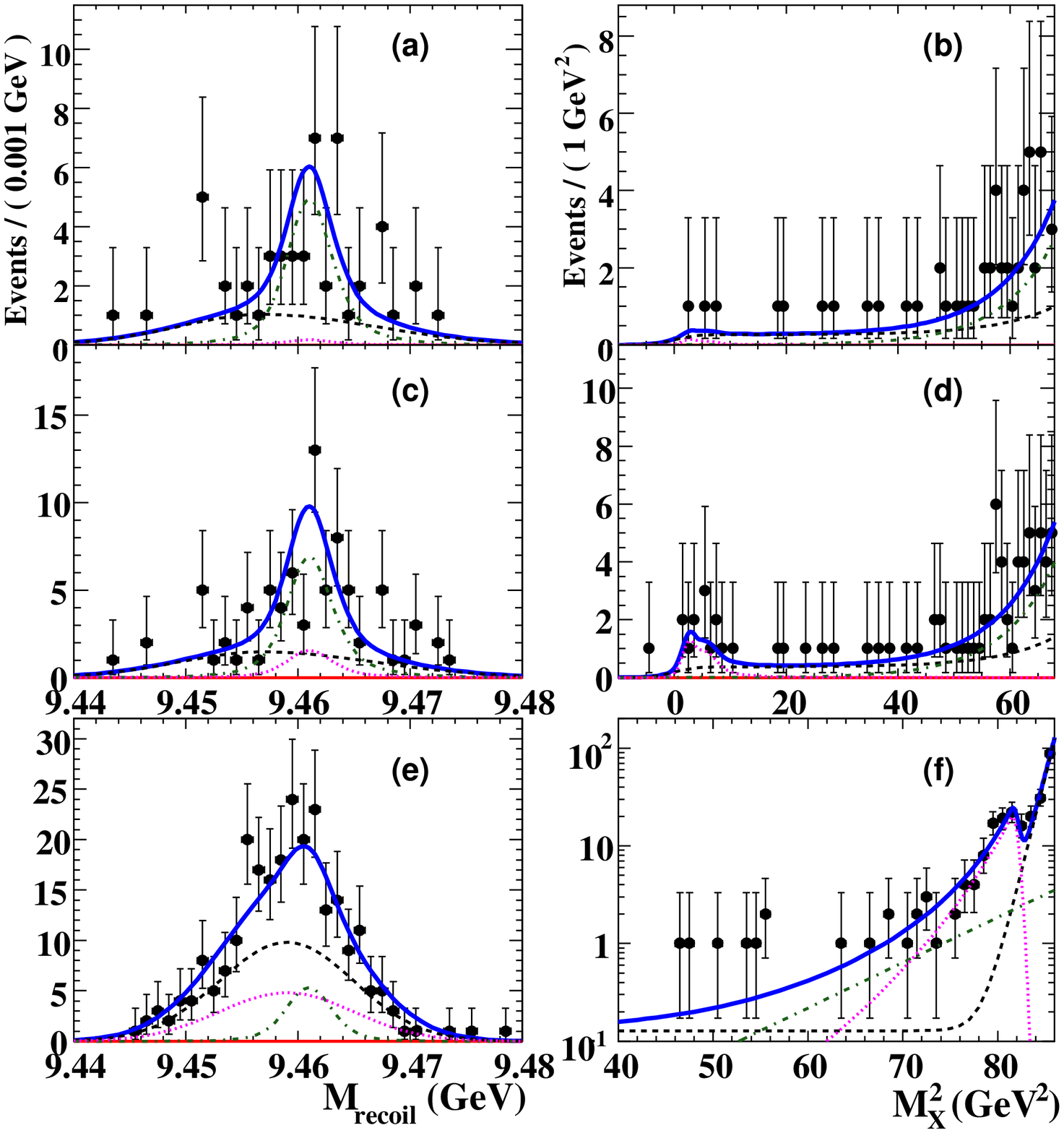,width=0.5\textwidth}
\ec
\vspace{-1.5\baselineskip}
\caption{Projection plots from the fit 
  with $N_\mathrm{sig}=0$ onto (a,c,e)
  \recMass\ and (b,d,f) \missMass. (a,b): low-mass region with IFR
  veto; (c,d): low-mass region without IFR veto; (e,f): high-mass
  region. 
  Overlaid is the fit with $N_\mathrm{sig}=0$ (solid blue line),
  continuum background (black dashed line), radiative
  leptonic \Y1S decays (green dash-dotted line), and (c,d) radiative
  hadronic \Y1S decays or (e,f) $\eta'$ background (magenta dotted line). 
}
\label{fig:unblindZero}
\end{figure}

As a first step in the likelihood scan, we perform fits to
the low-mass and high-mass regions with $N_\mathrm{sig}=0$. The free
parameters in the fit are $N_\mathrm{cont}$, $N_\mathrm{lept}$, and
$N_\mathrm{hadr}$ (low-mass region), and $N_\mathrm{cont}$,
$N_\mathrm{lept}$, and $N_\mathrm{\eta'}$ (high-mass region). The
results of the fits are shown in Fig.~\ref{fig:unblindZero}. We observe
no significant deviations from the background-only hypothesis. We find
$N_\mathrm{hadr}=8.7^{+4.0}_{-3.3}\pm0.8$ (without IFR veto) with a
significance of $3.5\sigma$, including systematic uncertainties. 

%%%%%%%%%%%%%%%%%%%%%%%%%%%%%%%%%%%%%%%%
\begin{figure}
\bc
\epsfig{file=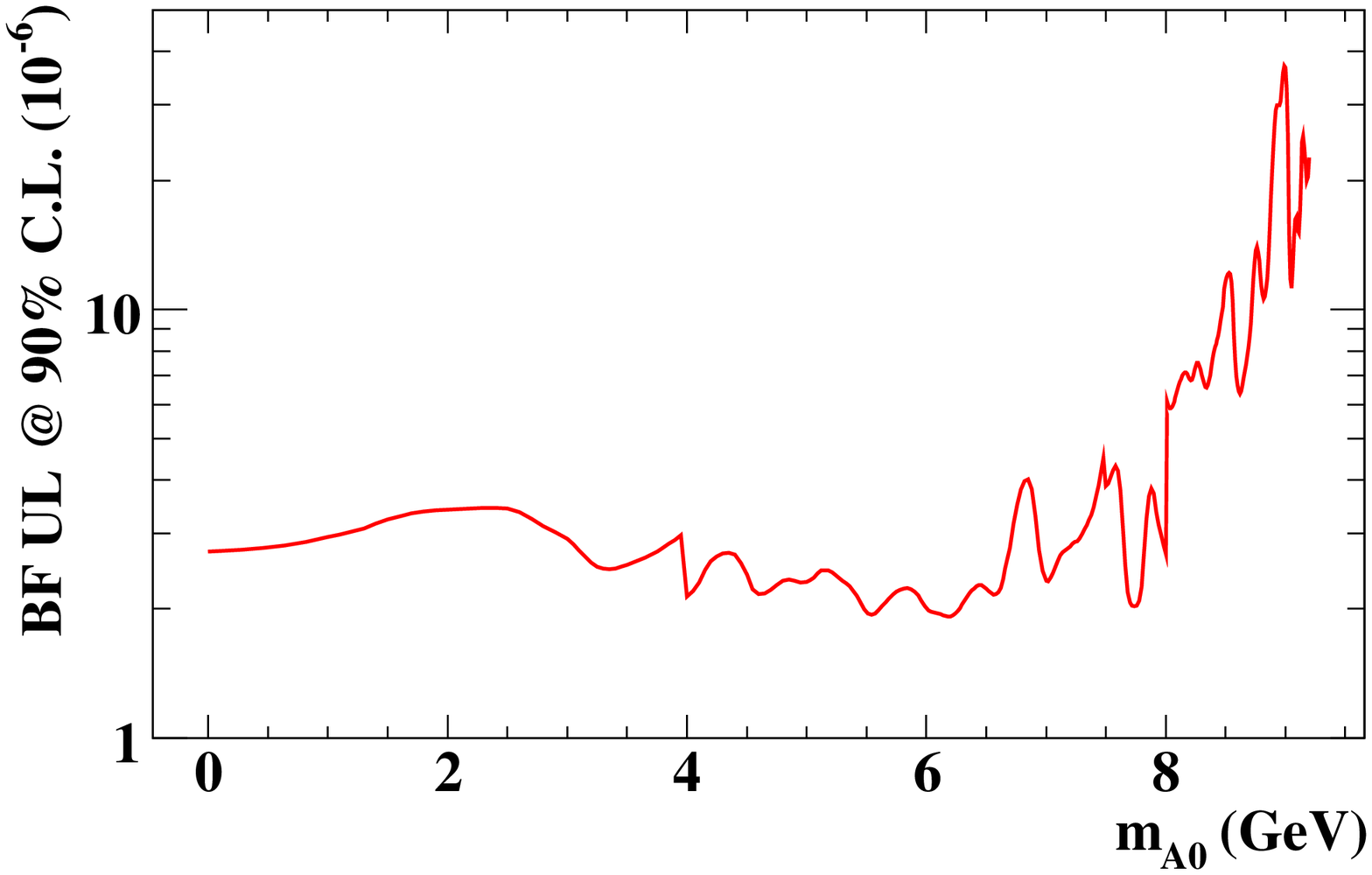,width=0.4\textwidth}
\ec
\vspace{-2\baselineskip}
\caption{90\% C.L. upper limits for 
  $\BR(\Y1S\to\gamma\cpoddhiggs)\times\BR(\cpoddhiggs\to\invisible)$.
}
\label{fig:unblindScanComb}
\end{figure}
\begin{figure}
\bc
\epsfig{file=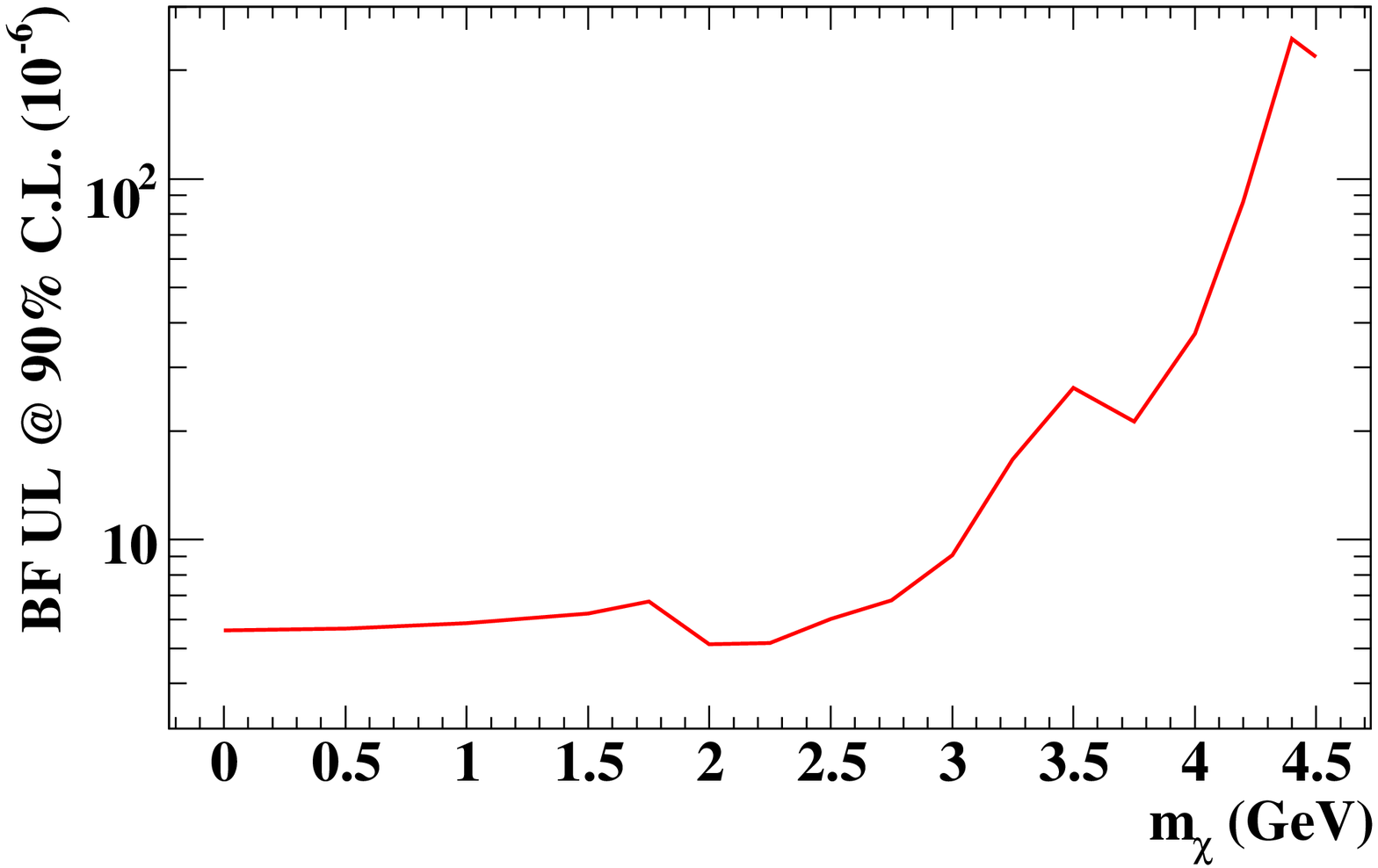,width=0.4\textwidth}
\ec
\vspace{-2\baselineskip}
\caption{90\% C.L. upper limits for $\BR(\Y1S\to\gamma\dm\dmbar)$.
}
\label{fig:unblindScanCombDM}
\end{figure}
%
%%%%%%%%%%%%%%%%%%%%%%%%%%%%%%%%%%%%%%%%

%%%%%%%%%%%%%%%%%%%%%%%%%%%%%%
% RESULTS AND CONCLUSIONS
%%%%%%%%%%%%%%%%%%%%%%%%%%%%%%

We then proceed to perform the likelihood scans as a function of
$N_\mathrm{sig}$ in steps of \higgsmass\ and \dmmass. In the scan, the
contribution of radiative hadronic background is fixed to the
expectation 
$N_\mathrm{hadr}=1.02\pm0.14$ for $\higgsmass<4$~GeV
($\dmmass<2$~GeV) where the IFR veto is applied, and
to $N_\mathrm{hadr}=6.6\pm1.1$ for fits in $4\le\higgsmass\le8$~GeV 
($2\le\dmmass<4$~GeV) range. 
We do not observe a significant excess of events above the background, 
and set upper limits on 
$\BR(\Y1S\to\gamma\cpoddhiggs)\times\BR(\cpoddhiggs\to\invisible)$
(Fig.~\ref{fig:unblindScanComb})
and $\BR(\Y1S\to\gamma\dm\dmbar)$
(Fig.~\ref{fig:unblindScanCombDM}). The limits are dominated by
statistical uncertainties. 
The largest statistical fluctuation, $2.0\sigma$, is observed at
$\higgsmass=7.58$\,\gev~\cite{EPAPS}; we estimate the probability to see such a
fluctuation {\em anywhere\/} in our dataset to be over 30\%. 

In summary, we find no evidence for the single-photon decays
$\Y1S\to\gamma+\invisible$, and set 90\% C.L. upper limits
on 
$\BR(\Y1S\to\gamma\cpoddhiggs)\times\BR(\cpoddhiggs\to\invisible)$
in the range (1.9--4.5)$\times10^{-6}$ for
$0\le\higgsmass\le8.0\,\gev$,  (2.7--37)$\times10^{-6}$ for
$8\le\higgsmass\le9.2\,\gev$,
and scalar \cpoddhiggs. 
We limit $\BR(\Y1S\to\gamma\dm\dmbar)$ in the range
(0.5--24)$\times10^{-5}$ at 90\% C.L. for
$0\le\dmmass\le4.5\,\gev$, assuming the phase-space distribution of
photons in this final state.
Our results improve the existing limits by an order of magnitude or
more, and significantly constrain~\cite{EPAPS} light Higgs
boson~\cite{Dermisek:2006py} and 
light dark matter~\cite{Yeghiyan:2009} models.

%%%%%%%%%%%%%%%%%%%%%%%%%%%%%%
% ACKNOWLEDGMENTS
%%%%%%%%%%%%%%%%%%%%%%%%%%%%%%

% Standard acknowledgments paragraph; must always be included.
We are grateful for the excellent luminosity and machine conditions
provided by our \pep2\ colleagues, 
and for the substantial dedicated effort from
the computing organizations that support \babar.
The collaborating institutions wish to thank 
SLAC for its support and kind hospitality. 
This work is supported by
DOE
and NSF (USA),
NSERC (Canada),
CEA and
CNRS-IN2P3
(France),
BMBF and DFG
(Germany),
INFN (Italy),
FOM (The Netherlands),
NFR (Norway),
MES (Russia),
MICIIN (Spain),
STFC (United Kingdom). 
Individuals have received support from the
Marie Curie EIF (European Union),
the A.~P.~Sloan Foundation (USA)
and the Binational Science Foundation (USA-Israel).

% Specific acknowledgments for this paper; remove if not needed.

\onecolumngrid
\newpage

\section{Appendix: EPAPS Material}

The following includes supplementary material for the Electronic
Physics Auxiliary Publication Service.

\begin{figure}[h!]
\begin{center}
\subfigure[]{\epsfig{file=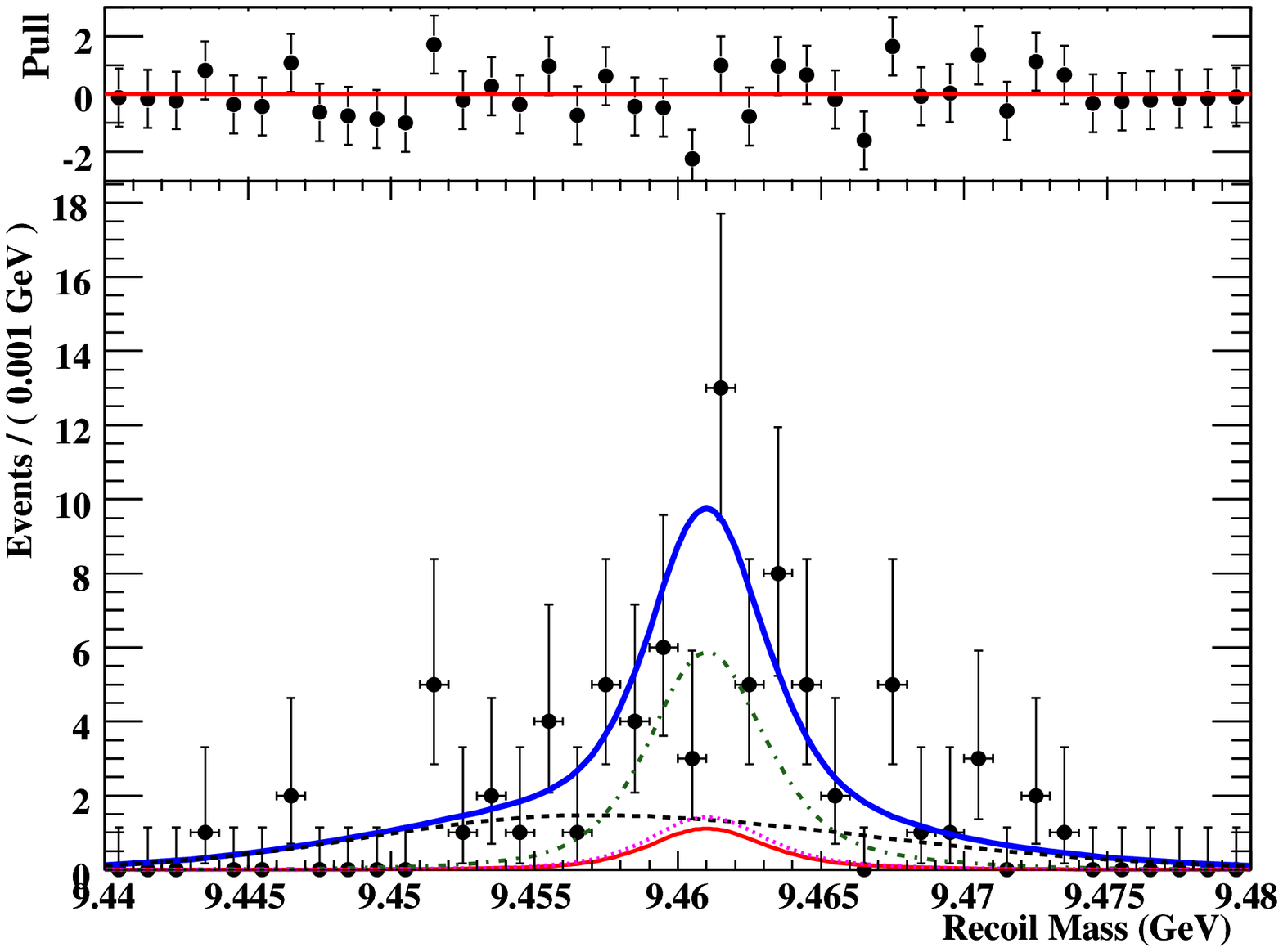,width=0.45\textwidth}}\hspace{0.1 in}
\subfigure[]{\epsfig{file=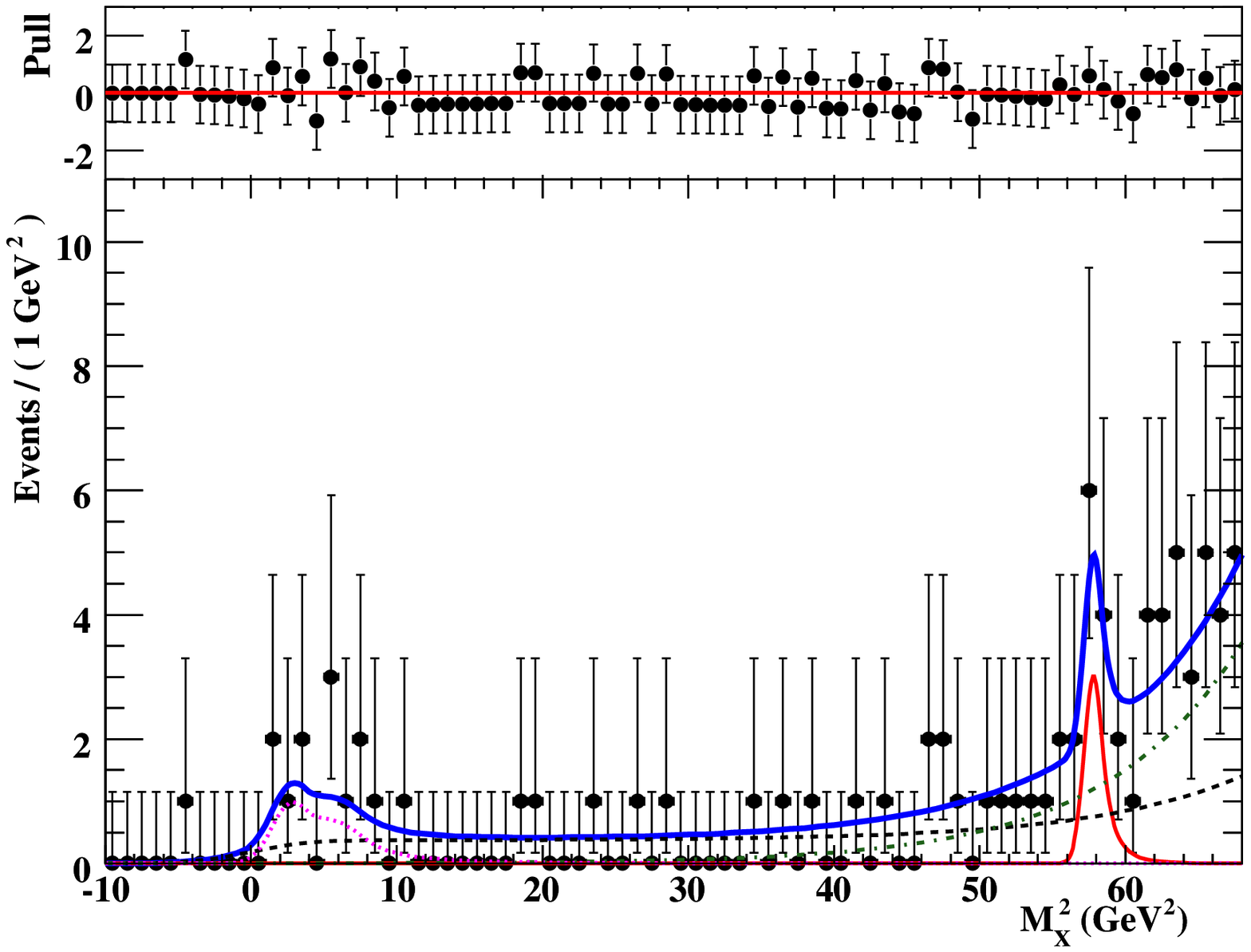,width=0.45\textwidth}}
\end{center}
\caption{Projection plots from the fit 
  with $\higgsmass=7.58$\,\gev (the most significant deviation from zero) to (a)
  \recMass\ and (b) \missMass. 
  Overlaid is the fit (solid blue line), signal contribution (solid
  red line), 
  continuum background (black dashed line), radiative
  leptonic \Y1S decays (green dash-dotted line), and radiative
  hadronic \Y1S decays (magenta dotted line). The top plot show
  residuals in each bin, normalized by the bin error. 
  The fit corresponds to 
  $\BR(\Y1S\to\gamma\cpoddhiggs)\times\BR(\cpoddhiggs\to\invisible)=(3.2^{+2.2}_{-1.8}\pm1.0)\times10^{-6}$,
  where the first uncertainty is statistical and the second is
  systematic, and statistical significance of $2.0\sigma$. The
  probability to observe such a
  fluctuation {\em anywhere\/} in our dataset is over 30\%.
}
\label{fig:unblind7.58}
\end{figure}
\begin{figure}[h!]
\begin{center}
\epsfig{file=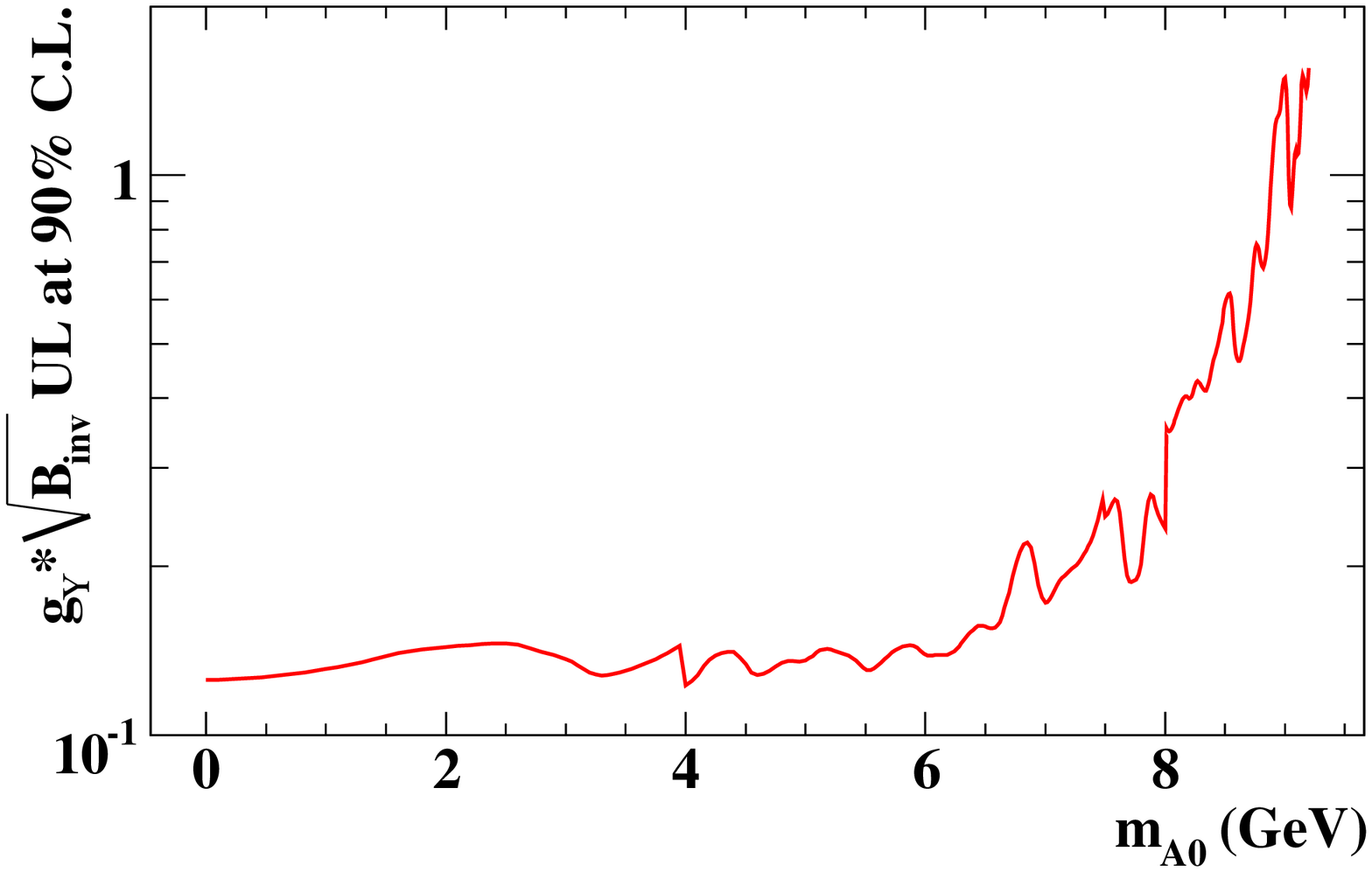,height=3in}
\end{center}
\caption{Upper limits on the product
  $g_{\Upsilon}\times\sqrt{\mathcal{B}(\cpoddhiggs\to\mathrm{invisible})}$
  at 90\% C.L. as a function of \higgsmass. The parameter
  $g_{\Upsilon}$ 
  is an effective coupling of the CP-odd Higgs
  \cpoddhiggs\ to bound state $\Upsilon(1S)$; in NMSSM, 
  $g_{\Upsilon}=\tan\beta\cos\theta F_{\Upsilon}$, where $\cos\theta$
  is the fraction of non-singlet component in \cpoddhiggs, $\tan\beta$
  is the ratio of Higgs vacuum expectation values, and $F_{\Upsilon}$
  is the effective form-factor (including the QCD and QED
  corrections). 
  The theoretically preferred region in NMSSM~\cite{Dermisek:2006py} is
  $g_{\Upsilon}>1$. }
\label{fig:coupling}
\end{figure}


\begin{thebibliography}{99}

\bibitem{ref:DM}
E.\ Komatsu \etal\ (WMAP Collaboration),
Astrophys.\ J.\ Suppl.\ \textbf{180}, 330 (2009);
M.\ Tegmark \etal\ (SDSS Collaboration), 
\jprd{74}, 123507 (2006).

\bibitem{ref:DMreview}
G.\ Bertone, D.\ Hooper, and J.\ Silk, 
Phys. Rep. \textbf{405}, 279 (2005).

\bibitem{PDBook}
K.\ Nakamura \etal, (Particle Data Group), Journal of Physics G 
{\bf 37}, 075021 (2010). 

\bibitem{Gunion:2005rw}
 J.\ F.\ Gunion, D.\ Hooper, and B.\ McElrath,
  \jprd{73}, 015011 (2006).

\bibitem{INTEGRAL}
P.\ Jean \etal, Astron. Astrophys. \textbf{407}, L55 (2003); 
J.\ Knodlseder \etal, Astron.\ Astrophys. \textbf{411}, L457 (2003).

\bibitem{DAMA}
R.\ Bernabei \etal\ (DAMA Collaboration), Eur.\ Phys J.\ \textbf{C56},
333 (2008). 

\bibitem{BAD2009prl}
B.\ Aubert \etal\ (\babar\ Collaboration),
\jprl{103}, 251801 (2009). 

\bibitem{Yeghiyan:2009}
G.\ Yeghiyan, \jprd{80}, 115019 (2009).

\bibitem{McElrath:2007sa}
R.\ McElrath, preprint arXiv:0712.0016 [hep-ph] (2007);
P.\ Fayet, \jprd{81}, 054025 (2010). 

\bibitem{Wilczek:1977zn}
F.\ Wilczek, \jprl{39}, 1304 (1977). 

\bibitem{Dermisek:2005ar}
R.\ Dermisek and J.~F.\ Gunion, \jprl{95}, 041801 (2005). 

\bibitem{LisantiWacker2009}
M.\ Lisanti and J.~G.\ Wacker, \jprd{79}, 115006 (2009).

\bibitem{Dermisek:2006py}
R.\ Dermisek, J.~F.\ Gunion, and B.\ McElrath, \jprd{76}, 051105 (2007);
R.\ Dermisek and J.~F.\ Gunion, \jprd{81}, 075003 (2010).

\bibitem{ShrockSuzuki1982}
R.~E.\ Shrock and M.\ Suzuki, \jpl{B110}, 250 (1982).

\bibitem{CLEO:1994ch}
R.\ Balest \etal\ (CLEO Collaboration), \jprd{51}, 2053 (1995). 

\bibitem{ref:Lozano}
E.\ Fullana and M.~A.\ Sanchis-Lozano, \plb{653}, 67 (2007). 

% The NIM detector performance paper
\bibitem{detector}
B.\ Aubert \etal\ (\babar\ Collaboration),
\nima{479}, {1} (2002).

\bibitem{LST}
W.\ Menges, IEEE Nucl. Sci. Symp. Conf. Rec. \textbf{5}, 1470 (2006).

\bibitem{CMfootnote}
Henceforth, $^{*}$ marks CM quantities.

\bibitem{TMVA}
A.\ H\"ocker \etal, preprint physics/0703039,
PoS ACAT, 040 (2007); \texttt{http://tmva.sourceforge.net/}.

\bibitem{CLEOgammah+h-}
S.~B.\ Athar \etal\ (CLEO Collaboration), \jprd{73}, 032001 (2006).

\bibitem{Punzi}
G.\ Punzi, preprint physics/0308063 (2003).

\bibitem{geant}
S. Agostinelli \etal\ (\textsc{Geant4} Collaboration),  
Nucl. Instrum. Methods Phys. Res.,
Sect. A {\bf 506}, 250 (2003).

\bibitem{EvtGen}
D.\ Lange, Nucl. Instrum. Methods Phys. Res.,
Sect. A {\bf 462}, 152 (2001).

\bibitem{ref:CBshape}
M.~J.~Oreglia, Ph.D Thesis, report SLAC-236 (1980), Appendix D.

\bibitem{EPAPS}
Additional plots are available in the Appendix. 


\end{thebibliography}
\end{document}